\documentstyle[emulateapj,psfig]{article}

\newcommand{\be}{\begin{equation}}
\newcommand{\ee}{\end{equation}}
\newcommand{\ba}{\begin{eqnarray}}
\newcommand{\ea}{\end{eqnarray}}
\newcommand{\siml}{\lower4pt \hbox{$\buildrel < \over \sim$}}
\newcommand{\simg}{\lower4pt \hbox{$\buildrel > \over \sim$}}

\begin{document}

\title{On the Kinetic Energy and Radiative Efficiency of Gamma-Ray Bursts}
\author{Nicole M. Lloyd-Ronning$^1$, Bing Zhang$^2$}
\affil{$^{1}$ Los Alamos National Lab,, MS D436, Los Alamos NM, 
87544; ronning@lanl.gov \\ 
$^{2}$Department of Astronomy \& Astrophysics, Pennsylvania
State University, University Park, PA 16803;
bzhang@astro.psu.edu\\}

\begin{abstract} 
Using measured X-ray luminosities to 17 Gamma-Ray
Bursts (GRBs) during the afterglow phase and accounting for radiative 
losses, we calculate the kinetic
energy of these bursts and investigate its relation to other GRB
properties.  We then use the observed radiated energy during the
prompt phase to determine the radiative efficiency of these bursts,
and explore how the efficiency relates to other GRB observables.  We
find that the kinetic energy in the afterglow phase is directly
correlated with the radiated energy, total energy as well as possibly
the jet opening angle and spectral peak energy.  
More importantly, we find the intriguing fact that the efficiency
is correlated with the 
radiated energy, and mildly with the total energy, jet opening angle
and spectral peak energy. XRF020903 also seems to follow the trends we 
find for our GRB sample. We discuss the implications of these
results for the GRB  radiation and jet models.
\end{abstract}

\section{Introduction}
  In the past several years, there has been a significant increase in 
our understanding of the energetics of
 Gamma-Ray Bursts (GRBs).  By measuring the fluence during the prompt 
phase (usually when most of the energy is emitted), and determining the 
burst redshift and degree to which the outflow is beamed from afterglow,
 one can make an estimate of the radiated 
energy of the burst.  Frail et al. (2001) showed that in fact most GRBs 
with measured redshifts have  radiated energies distributed very
narrowly around the value $5 \times 10^{50}$ erg. This conclusion is 
confirmed by Bloom, Frail
\& Kulkarni (2003) with a larger jet data sample.  
Understanding this energy output is of course crucial in helping constrain 
progenitor models. 
 However, an often
overlooked point in discussing GRB 
energetics is that the radiated energy is not necessarily the {\em total} energy of 
the burst.  
The radiated energy is some unknown fraction of the initial kinetic energy 
of the outflow - i.e. the kinetic energy is converted to radiation via 
some shock (or other) interaction(s) in the burst ejecta.  It is this 
kinetic energy or, more precisely, the {\em efficiency} of converting 
kinetic energy to radiation that must be considered when discussing GRB 
energetics. 
This efficiency may help us better understand by what 
mechanism the kinetic energy becomes radiation. It may also play an 
important role in understanding 
so-called ``X-ray Flashes'' (XRFs; GRBs with lower spectral peak energy 
and flux) - i.e.  XRFs may well be GRBs with much smaller emission 
efficiency.  To assess the kinetic energy, the most adequate method
is through broadband afterglow fits (e.g. Panaitescu \& Kumar 2001),
while another convenient method is to study the X-ray afterglow data alone
(Freedman \& Waxman 2001). Using both methods, it has been found that
the kinetic energy - corrected for the degree of beaming of
the outflow -  is also 
narrowly distributed (Panaitescu \& Kumar 2001; Piran et al. 2001, under the assumption 
that there is a correlation between X-ray luminosity and beaming corrections ). The result (as 
well as the validity of the Piran assumption)
is confirmed by Berger, Kulkarni \& Frail (2003) with a larger X-ray
afterglow data sample. { Although these works are the first to present calculations
of burst kinetic energies, to our knowledge
there are  no detailed 
studies of correlations between kinetic energy or GRB efficiency with other burst properties.}

  With these points in mind, we  derive the absolute values of GRB 
kinetic energy for a large data sample, and investigate the behavior
of the kinetic 
energy and efficiency of GRBs, in relation to other burst properties (such 
as radiated energy, total energy, characteristic jet angle, and spectral peak 
energy).  The paper is organized as follows:  In \S 2, we describe how to 
compute the kinetic energy and efficiency of GRBs from the observed 
{ X-ray} afterglow data.  We also discuss the basic assumptions and
caveats with  
our method. In \S 3, we present our results, including the correlation of 
kinetic energy and efficiency with other burst properties. Our main 
results shown that both kinetic energy and efficiency are correlated with 
radiated energy and to some extent, total energy.  There are also 
suggestions of relations with the jet opening (or viewing) angle and the
spectral peak energy, although additional data are needed to confirm this. 
In \S 4, we 
discuss the physical implications of these results, and in \S 5, present 
our summary and conclusions. 
  
\section{Computing GRB Efficiency}
 
  We define the GRB efficiency as 
\ba
 \zeta \equiv E_{\gamma}/(E_{k} + E_{\gamma}) 
\ea 
where $E_{\gamma}$ is the
radiated energy in the prompt phase and $E_{k}$ is the burst kinetic
energy remaining during the afterglow phase (see below). In what 
follows, all energies will refer to the isotropic equivalent
energies. {Since both energy components are corrected the same way 
when the beaming factor included, the GRB efficiency does not depend
on the unknown beaming configuration, and is therefore jet model
independent.} The radiated 
energy is obtainable through direct measurement, and we take ours from 
Table 2 of Bloom, Frail and Kulkarni (2003).  The kinetic energy can be 
readily calculated from the X-ray luminosity during the afterglow phase, in 
the context of the standard afterglow model (M\'esz\'aros \& Rees
1997; Sari, Piran \& Narayan 1998).
In this model, the specific flux at a particular frequency depends on,
among other parameters, the total kinetic energy in the fireball.
 In particular, since at a late afterglow epoch the X-ray band is 
above the cooling frequency, the X-ray luminosity is only 
sensitively dependent on $E_k$ and $\epsilon_e$ \footnote{When a radiative 
correction is taken into account it is no
longer sensitive to $epsilon_e$; see below for more discussion. } (the 
electron
equipartition parameter), so that the X-ray luminosity at a certain
afterglow epoch is a sensitive probe for $E_k$ (Freedman \& Waxman
2001). 

 Using the standard afterglow model 
(e.g. Hurley, Sari \& Djorgovski 2002), we can
calculate $\nu_m$, the characteristic frequency corresponding to the
minimum electron energy, 
$\nu_c$, the synchrotron ``cooling'' frequency, and $F_{\nu,m}$, the
{ peak synchrotron flux}, at 10 hr after the burst trigger.
\ba
\nu_m & = & 1.15\times 10^{15}~{\rm Hz}\left(\frac{p-2}{p-1}\right)^{2}
\left(\frac{1+z}{2}\right)^{1/2} \nonumber \\
& \times & E_{52}^{1/2} \epsilon_{e,-1}^2
\epsilon_{B,-2}^{1/2} t_{10h}^{-3/2}, \\
\nu_c & = & 9.9\times 10^{15}~{\rm Hz}
\left(\frac{1+z}{2}\right)^{-1/2} \nonumber \\
& \times & E_{52}^{-1/2} \epsilon_{B,-2}^{-3/2}
n^{-1} t_{10h}^{-1/2}, \\
F_{\nu,m} & = & 4.0 \times 10^{-26}~{\rm erg~s^{-1}~cm^{-2}~Hz^{-1}}
\left(\frac{1+z}{2}\right) \nonumber \\
& \times & E_{52} \epsilon_{B,-2}^{1/2} n^{1/2} D_{L,28}^{-2}.
\label{ag}
\ea
where $p$ is the electron particle spectrum index, $z$ is the burst redshift,
$n$ is the ambient density assumed to be constant here, 
$E_{52}=E_{k}/10^{52}~{\rm erg}$, $\epsilon_{e,-1}=\epsilon_e
/ 0.1$ is the fraction of energy in the electrons
, $\epsilon_{B,-2}=\epsilon_B/0.01$ is the fraction of energy in the magnetic
field, $t_{10h}=t/10~{\rm hr}$ is the time of observation,
and $D_{L,28}=D_L/10^{28}~{\rm cm}$, where $D_L$ is the luminosity
distance.

We consider the X-ray band with typical frequency $\nu=10^{18}~{\rm Hz}
\nu_{18}$ (which corresponds to $\sim 5$ keV, the mid-energy for the
current X-ray detectors such Chandra or XMM-Newton). The specific flux at
this frequency at 10 hours after the burst trigger is:
\ba
F_{\nu,x}({\rm 10 hr}) & = & F_{\nu,m}\nu_m^{(p-1)/2} \nu_c^{1/2}
\nu^{-p/2} \nonumber \\
& = & 8.0\times 10^{-30}~{\rm erg~s^{-1}~cm^{-2}~Hz^{-1}} \nonumber \\
& \times & \left(\frac{1+z}{2}\right)^{(p+2)/4} E_{52}^{(p+2)/4}
\epsilon_{e,-1}^{p-1} \epsilon_{B,-2}^{(p-2)/4} \nonumber \\
& \times & t_{10h}^{(2-3p)/4} \nu_{18}^{-p/2} D_{L,28}^{-2}.
\ea
To get the numerical value of the coefficient, $p=2.2$ has been
adopted. The power indices for various parameters are consistent with
Freedman \& Waxman (2001). The ``isotropic'' X-ray afterglow
luminosity at 10 hour is then
\ba
L_x({\rm 10 hr}) & = & 4 \pi D_L^2 \nu F_{\nu,x}(10h) \nonumber \\
& = & 1.0\times 10^{46}~{\rm erg~s^{-1}}
\left(\frac{1+z}{2}\right)^{(p+2)/4} \nonumber \\
& \times & E_{52}^{(p+2)/4}
\epsilon_{e,-1}^{p-1} \epsilon_{B,-2}^{(p-2)/4} t_{10h}^{(2-3p)/4}
\nu_{18}^{(2-p)/2}.
\ea
We  finally get
 \ba
 E_{k} & = & 10^{52}~{\rm erg}
{\cal R}\left(\frac{L_x({\rm 10 hr})} { 10^{46} ~{\rm
erg~s^{-1}}}\right)^{4/(p+2)} \left(\frac{1+z}{2}\right)^{-1}
\nonumber \\
& \times & 
\epsilon_{e,-1}^{4(1-p)/(2+p)} \epsilon_{B,-2}^{(2-p)/(2+p)}
t_{10h}^{(3p-2)/(p+2)} \nu_{18}^{2(p-2)/(p+2)}.
 \ea 
where $\cal{R}$ is a
factor that accounts for radiative losses during the first 10 hours 
following
the prompt phase (Sari 1997), 
\be
{\cal R}= \left[\frac{t(10h)}{t(prompt)}\right]^{(17/16)\epsilon_e}.
 \ee

Taking $t_{10h}=1$ and $\nu_{18}=1$, we can use the X-ray luminosity
measurements at 10 hours from Berger, Kulkarni \& Frail (2003; their 
Table 2) to calculate $E_{k}$ at this time during the afterglow
(without the ${\cal R}$ correction).
However, this kinetic energy is the amount left 
at 10 hours after the prompt phase.  Including the correction factor
${\cal R}$ for radiative losses during these initial 10 hours, we can
estimate the value of the 
kinetic energy {\em at the end of the prompt phase} $\sim 50s$.  A list 
of our energies and efficiencies is presented in Table 1.
We note that
our efficiencies are relatively high (between 0.4 and 1.0)
which challenges the simplest internal shock model (Kumar 1999,
Panaitescu, Spada, \& Meszaros 1999),
 although it has been shown that such efficiencies are indeed
achievable in an internal shock scenario (Beloborodov 2000, Kobayashi \&
Sari 2001).

Our error bars on $E_{\gamma}$ are taken directly from measurement
error listed in Table 2 of Bloom, Frail and Kulkarni (2003).  For
$E_{k}$, we compute the error bar from the measurement error of the
X-ray luminosity listed in Table 2 of Berger, Kulkarni, and Frail
(2003), and assume that all other parameters are known constants
(indeed an important assumption as described in the next subsection).
Finally, with both an error bar on $E_{\gamma}$ and $E_{k}$ we can
compute the error bar on the efficiency $\zeta$ by the standard
technique of error propagation (see, e.g, Bevingtion \& Robinson,
2002).

Only one XRF (XRF 020903) has been studied to infer $E_k$
(Soderberg et al. 2004). In Table 1, we include this event although 
the error bars on the kinetic and radiated energies are not
available.  Because of only 
one data point,
 for all the correlations discussed below, this XRF event is
excluded. The data point is however included in the figures when relevant.

 \subsection{Caveats}
 
   The absolute value of the kinetic energy of course plays a big role in
the final conclusions we draw from our analysis.  This expression depends
on the parameters $\epsilon_e, \epsilon_{B}$, and $p$, but - as one can
see from equation 7 above - appears to be most sensitive to $\epsilon_e$.  
Without the radiative correction $\cal{R}$, the kinetic energy decreases
(increases) as $\epsilon_{e}$ increases (decreases) as $E_{k} \propto
\epsilon_{e}^{\sim -1}$.  Correcting for radiative losses, however,
offsets this sensitivity to some degree since the exponent of the
$\cal{R}$ factor depends on $\epsilon_{e}$.  For the range of $\epsilon_e$
most commonly suggested by the data (see e.g., Panaitescu \& Kumar
2001; Yost et al. 2003)
the kinetic energy turns out to be fairly insensitive to $\epsilon_e$.
Meanwhile, increasing the exponent $p$ causes the kinetic energy to
increase and hence the efficiency to decrease.  The dependence on
$\epsilon_B$, as mentioned above is minimal. We also point
out that  strictly speaking the efficiency we define here
(i.e. $\zeta=E_{\gamma}/(E_{\gamma}+E_{k}$)) is only an ``apparent
efficiency''. It is the real efficiency only when there is no
additional kinetic energy injection in the afterglow phase (before 10
hour). This caveat has important implications for the discussions
about the jet models in \S4.

  In what follows below, we present results for $\epsilon_e=0.3$, 
$\epsilon_{B}=0.01$ and $p=2.2$.  These are typical values obtained from 
multiwavelengths to the afterglow spectra  (Panaitescu \& Kumar 2001;
Yost et al. 2003). For the often quoted ``standard'' value of
$\epsilon_e=0.1$, we get {\em  quantitatively}
the same results within the error bars.  
Our results are {\em qualitatively} the same for $\epsilon_e$ in the range 
of $  
0.01-0.3 $, $p$ in the range $2.2 -2.5 $ and $\epsilon_B$ in the range 
$0.001 - 0.1 $.

\section{Results}
  We have analyzed the relationship between $E_{k}$ and $E_{\gamma}$, 
total energy $E_{tot}$, characteristic jet angle $\theta_{j}$ (as discussed
below this is the jet opening angle in the uniform jet model and 
essentially the observer's
viewing angle with respect to the jet axis in the quasi-universal jet model), \& 
spectral peak energy $E_{p}$, as well as $\zeta$ with these variables.  
   In the text, we present all of our analysis, but
  only show the figures for the variables that exhibit either
statistically significant
  correlations or have implications in interpreting the physics of the 
outflow. Note that we consider a
correlation
  ``statistically significant'' if it is $\ga 3\sigma$ according to a standard Kendell's
  $\tau$ test (e.g., Press et al., 1994).
   We also point out two important relations which will play a role in the 
   interpretation of our results below. These 
  are the so-called ``Frail'' and ``Amati'' relations.  The Frail relation (Frail et al., 2001)
  shows a distinct {\em negative} correlation between the isotropic
radiated energy $E_{\gamma}$
  and the jet opening angle in a uniform jet model $\theta_{j}$, such that
  $E_{\gamma} \propto \theta_{j}^{-2}$.  The consequence of this is that
the geometry corrected
  emitted energy turns out to be approximately constant (Frail et al.,
2001, Bloom, Frail \&
  Kulkarni, 2003).  This can be interpreted in two ways.  In the ``uniform jet model'' (Frail et al., 2001),
   this implies
  that the same energy is collimated into different opening angles $\theta_{j}$ 
  varying from about 1 to 30 degrees.
  In the ``quasi-universal jet model'' (Rossi, Lazzati \& Rees, 2002;
Zhang \& Meszaros 2002; Lloyd-Ronning, Dai \& Zhang 2004; Zhang et
al. 2004), all GRBs have approximately the same jet structure with
  an energy that varies (decreases) as a function of angle
$\theta_{j}$ from the jet axis.
  We will return to these models for the GRB jet structure
  in the discussion.
  The Amati relation is a relation between the isotropic emitted energy
  and the spectral peak energy such that $E_{\gamma} \propto E_{p}^{2}$
(Lloyd, Petrosian \& Mallozzi
  2000; Amati et al., 2001; Lamb, Donaghy \& Graziani 2004, Liang, Dai \& 
Wu, 2004)
  This relation may be providing a key to the GRB emission mechanism
(Lloyd, Petrosian 
  \& Mallozzi, 2000) and possibly the structure of the GRB jet.
  
\subsection{$E_{\gamma}-E_{k}$}
  A Kendell's $\tau$ test gives the correlation between
the radiated energy $E_{\gamma}$ and the kinetic energy $E_{k}$
to be approximately $3 \sigma$ with a functional form $E_{\gamma} \propto 
E_{k}^{1.0 \pm 0.4}$. These data are plotted in Figure 1. { We can
see that the XRF data point, with a kinetic energy of $1.0 \times 
10^{50}$ erg and a radiated energy of $1.1 \times 10^{49}$ erg 
qualitatively follows the similar dependence.} 

The tight correlation between the two components (except the two outliers)
is expected. It has been found that both components are essentially
constants when corrected by the geometric factor { (Berger et
al. 2003)}. This already hints that the two should be correlated.
This validates the practice of using $E_{\gamma}$ and $E_{k}$ 
interchangably as often done in the literature.  { We note that
Berger et al. have also pointed that some outliers in their sample with
apparently low kinetic energy also appear to have a low radiated energy, consistent
with the trend we find here.}

\subsection{$\zeta-E_{\gamma}$}
A Kendell's $\tau$ test gives the correlation between
the efficiency $\zeta$ and the radiated energy $E_{\gamma}$
to be approximately $3 \sigma$ with a functional form $\zeta \propto
E_{\gamma}^{.15 \pm 0.05}$.  These data are plotted in Figure 2.
{Notice that XRF 020903 also falls onto the same correlation.
This result is very intriguing.} Although for a single burst one
expects a higher radiated energy for a higher value of efficiency,
there is no { a priori reason} to expect such a correlation to hold
among bursts { without introducing any specific GRB models.}
{ In fact, if $E_\gamma$ and $E_k$ is linearly correlated, one
should expect $\zeta$ to be essentially a constant. Our result is
consistent with this expectation to first order, since the correlation 
index (0.15) is very shallow. Yet, a positive correlation 
is revealed, which hints that there is a high-order non-linear
dependence between $E_\gamma$ and $E_k$. Also such a positive
correlation potentially carries more information about GRB
radiation physics, and may hold the key to differentiate among GRB
models. We discuss this further in \S4.}

\subsection{$\zeta-E_{k}$}
The Kendell's $\tau$ test indicates that there is no correlation between
the efficiency $\zeta$ and the kinetic energy $E_{k}$.

\subsection{$E_{k}-E_{tot}$}
  A Kendell's $\tau$ test gives the correlation between
the kinetic energy $E_{k}$ and the total energy $E_{tot}$
to be approximately $3.6 \sigma$ with a functional form $E_{k} 
\propto
E_{tot}^{0.8 \pm 0.2}$. These data are plotted in Figure 3.
XRF 020903
also fits the trend we find in the GRB sample, as seen in Figure 3.
 Such a correlation is not surprising given the correlation between 
$E_{k}$ and $E_{\gamma}$ and the definition of $E_{tot}=E_{k}+E_{\gamma}$.

\subsection{$\zeta-E_{tot}$}
The Kendell's $\tau$ test indicates that there is no 
statistically significant correlation between
the efficiency $\zeta$ and the total energy $E_{tot}$.  
However, although 
the formal significance of the correlation is only $\sim 2\sigma$, the 
data plotted in Figure 4 suggest that a trend of increasing $\zeta$ for 
increasing $E_{tot}$. XRF020903 appears to also follow this relationship,
as evident in Figure 4.

\subsection{$E_{k}-\theta_{j}$}
  A Kendell's $\tau$ test indicates a very mild negative correlation 
between
the kinetic energy $E_{kin}$ and the jet characteristic angle $\theta_{j}$
at the $2.6 \sigma$ level, with a functional form $E_{k}
\propto
\theta_{j}^{-1.7 \pm 0.4}$. These data are plotted in Figure 5.
  This correlation can be understood as a result of the correlation 
between kinetic energy and radiated energy as well as the Frail relation 
which shows that radiated energy is inversely proportional to the jet 
opening angle $\theta_{j}$.  There was no break in the afterglow 
lightcurve observed for XRF 020903.  This could indicate that the 
characteristic jet angle is very large (which would imply a late or 
non-existent break in the afterglow light curve), which fits qualitatively
with the trend of decreasing kinetic energy for increasing jet angle, as 
we see here.

\subsection{$E_{k}-E_{p}$}
  A Kendell's $\tau$ test indicates a $2.2\sigma$ (i.e. no significant) 
  correlation
between
the kinetic energy $E_{k}$ and spectral peak energy $E_{p}$
although the eye suggests otherwise; assuming such a correlation the best 
fit functional form is $E_{k}
\propto
E_{p}^{1.5 \pm 0.5}$. 
 If this correlation is in fact present it can be understood as a 
consequence of the $E_{k}-E_{\gamma}$ correlation and the Amati relation 
which shows that $E_{\gamma}$ is correlated with $E_{p}$.  This implies 
that the kinetic energy should be correlated with $E_{p}$.  Taken 
at face value, the power-law index of the correlation
is also consistent with the Amati relation.  XRF 020903, with an $E_{p} 
\approx 5$ keV, seems to follow this trend.

\subsection{$\zeta-{\theta_{j}}$}
The Kendell's $\tau$ test indicates that there is no
statistically significant 
correlation between
the efficiency $\zeta$ and the jet opening angle $\theta_{j}$, although by 
eye this trend (decreasing efficiency for increasing jet angle) seems 
apparent, and could be explained as a consequence of the 
$\zeta-E_{\gamma}$ correlation and the Frail relation. These data are 
plotted in Figure 6.

\subsection{$\zeta-E_{p}$}
A Kendell's $\tau$ test gives the correlation between
the efficiency $\zeta$ and the spectral peak energy $E_{p}$
to be approximately $2.5 \sigma$.  These data are ploted in Figure 7.  One 
can see a clear outlier at high efficiency and $E_{p} \approx 350$ keV.  
Eliminating this outlier, the correlation has a significance of $3 \sigma$ 
with a functional form $\zeta 
\propto E_{p}^{0.45 \pm 0.1}$.  Such a relation could be explained by the 
$\zeta-E_{\gamma}$ correlation and the Amati relation.
Again, XRF 020903 seems to follow this trend with its low efficiency 
($\zeta = 0.1$)
and $E_{p} = 5$ keV.\\

{ Although fits to afterglow spectra do give surprisingly similar values
  for $\epsilon_e$, $\epsilon_B$, and $p$ among different bursts, we of course do
  not expect that all bursts have exactly the same values for these
parameters (although simulations of particle acceleration in shocks do 
find consistently an electron index with a universal value of 2.2).
  As mentioned in \S 2, it turns out that the kinetic energy is fairly insensitive to
  the values of $\epsilon_B$ and $\epsilon_e$ when the radiative correction is
  taken into account.  We can, however, further explore the sensitivity of our results
  to the value of the electron index $p$, by using the measured X-ray afterglow spectral
  index (i.e. column 5 of Table 1 of Berger, Kulkarni \& Frail 2003) to
  estimate the value of $p$ for each burst.  
  For example, at the time of the afterglow this index is measured, we are in
  an adiabatic, slow-cooling fireball scenario (see, e.g. Sari, Piran, \& Narayan, 1998).  In
  this case we expect that the X-ray spectral index $\alpha_{X} = 3p/4 - 1/2$.  We can use
  this equation to solve for $p$ for each burst, and use this $p$ in our computation of the kinetic
  energy for the GRB.  It is important to point out that there is large uncertainty in value of $\alpha_X$
  and in fact most of the data from
  Berger, Kulkarni \& Frail are consistent with an electron index $p=2.2$.  
Nonetheless, we perform this exercise
  as an additional check on the robustness of our results. We find that
when using individual values of
  $p$ computed from the X-ray afterglow spectral index, all of the correlations remain with the same functional form
  as reported above,  although the siginificance is slightly reduced in the case of the $\zeta-E_{\gamma}$ correlation (from
  $3\sigma$ to $2.5\sigma$.  In the case of the $E_{k}$-$E_{tot}$ correlation, the significance {\em increases} from
  $3.6\sigma$ to $4\sigma$.}

\section{Discussion}

 We have investigated some correlations between the GRB kinetic
energy or efficiency and other observables  such as radiated energy,
spectral peak energy, characteristic jet  angle, etc. 
Although some of these correlations are expected from the Frail and
Amati correlations, we do discover some new correlations, which
 give us a further glimpse into the physics of GRBs.
 In particular the correlation between efficiency $\zeta$ and radiated energy
 $E_{\gamma}$ has the potential of differentiating between different models
 for the GRB prompt phase (note that this correlation was also indirectly 
suggested in Lamb, Donaghy \& Graziani, 2003). { For example, this 
relation is consistent with the internal shock model. 
 This is because $E_{\gamma}$ has been found to be correlated to the
burst variability parameter (a measure of the ``spikiness'' of the prompt
phase lightcurve, Fenimore \& Ramirez-Ruiz 2000; Reichart et
al. 2001). { In the internal shock model of Kobayashi \& Sari (2001), the
shells separate after each collision and then collide again with other
shells.  Even with the same Lorentz factor contrast, more shells lead to more collisions and therefore both
 a higher efficiency and variability.} If this hypothesis is true, we should see
higher efficiency for bursts with higher variability. Unfortunately,
there are currently not enough data to explore a $\zeta$-variability
relation. This may be tested with future missions such as Swift or
GLAST. Similarly, future more data are needed to verify whether there
is a significant $\zeta-E_{tot}$ correlation.
} 
 
 The $\zeta-E_{\gamma}$ correlation also plays an important role in
interpreting  the other correlations found in \S 3 above, when
combined with previously determined relations ($E_{\gamma} \propto 
\theta_{j}^{-2}$ and $E_{\gamma} \propto E_{p}^{2}$).  For example,
the relationship between $\zeta$ and $E_{p}$ is a consequence of the
$\zeta-E_{\gamma}$ correlation and the Amati relation $E_{\gamma}
\propto E_{p}^{2}$.

{ At least for the GRB sample (excluding the XRF data point), we
have verified that using $E_k$ and $E_\gamma$ interchangeably, as done
previously in the studies of both the quasi-universal
structured jets (Zhang et al. 2004) and the uniform jets (Lamb et
al. 2004), is approximately valid. However, the discovered positive
$\zeta-E_\gamma$ correlation has profound implications for both jet
models. This correlation indicates that interchanging $E_k$ and
$E_\gamma$ is no longer a good approximation in XRFs ({ this point
has previously been made by Soderberg et al. (2003) for XRF020903}). In fact, $E_k$
of XRF 020903 is one order of magnitude larger than $E_\gamma$.
This fact is particularly fatal for the suggestion that GRBs have a
very narrow jet (Lamb et al. 2004) within the uniform jet model.
In that model, it is required that GRBs and XRFs all have a same
total energy. Since XRFs have a very small total prompt emission
energy, a narrow beam for GRBs is inferred (Lamb et al 2004). However,
our finding indicates that the total energy of XRFs is actually much
larger (if the efficiency is the real one). Keeping a constant energy,
the typical GRB beaming angle should be much larger than 1 degree. So
even using the arguments presented in Lamb et al. (2004), the GRB
beams are not narrow. 

Alternatively, as we have recently promoted (Zhang et al. 2004), GRBs
and XRFs may be unified within a framework of a quasi-universal
Gaussian-type jet model. In this model, we have also used $E_k$ and
$E_\gamma$ interchangeably for XRFs. However, the $E_k$ in this model
likely changes with time. Since the jet is structured with more energy
concentrated around the jet axis, there is a pole-to-equator energy
flow during the interaction of the jet with the ambient medium (Kumar
\& Granot 2003; Zhang et al. 2004). So in this model, the kinetic
energy measured at 10 hour after the burst could be much larger than
the initial value, especially when the line of sight is far away from
the jet axis (as is required to account for XRFs). Since our current
measured efficiency $\zeta$ is derived by extrapolating the inferred
kinetic energy measured at 10 hour back to the end of the prompt
emission without invoking any energy injection, the $\zeta-E_\gamma$
correlation could be simply an ``apparent'' correlation due to
the pole-to-equator energy flow. 

Of course, it could be the case that the $\zeta-E_\gamma$ correlation is
intrinsic. In this picture, XRFs are simply inefficient GRBs, but they no
longer necessarily share a ``standard'' total energy, and there is no
unified picture to incorporate both types of events.}

These relations and their physical implications can be further explored 
with the Swift satellite, set to launch in September, 2004. 
For example, if we can get a uniform sample of X-ray luminosities at
{ an early epoch (e.g. 1000 seconds or 1 hour)} as well as at later
times (say 10 hours), we may be able to test
whether there is additional injection of kinetic energy during the 
afterglow phase.  Swift may also allow the $\zeta$-variability relation to 
be explored in detail, { and eventually unveils the origin of the
$\zeta-E_{\gamma}$ correlation.}

 \section{Summary and Conclusions}
We have computed the kinetic energy and radiative efficiency of a
sample of 17 GRBs { and 1 XRF, accounting for radiative losses during 
the initial afterglow phase.}  We have found that both of these
GRB properties are related to a number of other GRB observables such
as radiated energy, total energy, jet characteristic angle, and
spectral peak energy.  We have shown that kinetic energy is directly
proportional to the radiated energy of the burst, as well as the total
energy and spectral peak energy.  The kinetic energy also decreases
with increasing jet characteristic (opening or viewing) angle. 
{ More importantly,} we
have also found that the efficiency is correlated with the radiated
energy of the burst, which appears to be consistent with an internal
shock picture for the GRB prompt phase. The efficiency also appears
to increase with increasing total energy and spectral peak energy, and
decreases with increasing jet angle. XRF 020903 (the only XRF for which we
are able to compute the kinetic energy and efficiency) seems to follow 
all of the trends we find for the 17 GRBs.

The relation between kinetic energy and radiated energy, efficiency
and radiated energy, combined with the previously known Frail and
Amati relations can explain the additional correlations we find in our
sample. { The results are compatiable with the 
quasi-universal model, although a large data sample for the GRB kinetic
energy derived at an earlier epoch (attainable from the Swift
obseratory) is essential to fully test this hypothesis. 
The results also strongly disfavor a ``narrow beam'' interpretation of
GRBs.} 

The relationship between efficiency and other GRB parameters is an
important one in elucidating GRB physics - we can further explore
these and other relations (such as efficiency and variability) and
potentially distinguish between various GRB { radiation and jet}
models with the launch of the Swift satellite in September, 2004.

\acknowledgments

We thank the anonymous referee for useful comments and 
suggestions.
B.Z. acknowledges NASA LTSA program for support.

\begin{deluxetable}{ccccccc}
\tablecolumns{4}
\tablewidth{0pc}
\tablecaption{\label{tab:effic}} 
\tablehead {
\colhead {GRB}        	&
\colhead {$E_{\gamma}/10^{52} {\rm erg}$} 	&
\colhead {$E_{k}/10^{52} {\rm erg}$}	&
\colhead {$\zeta$}		&
\colhead {}		
}
\startdata
970228 & $  1.42\pm   0.25$ & $  1.90\pm   0.21$ & $  0.43\pm   0.05$  \\ 
970508 & $  0.55\pm   0.06$ & $  0.99\pm   0.14$ & $  0.36\pm   0.04$  \\ 
970828 & $ 21.98\pm   2.40$ & $  4.06\pm   0.75$ & $  0.84\pm   0.03$  \\ 
971214 & $ 21.05\pm   2.58$ & $  8.48\pm   0.97$ & $  0.71\pm   0.03$  \\ 
980613 & $  0.54\pm   0.10$ & $  1.22\pm   0.38$ & $  0.30\pm   0.08$  \\ 
980703 & $  6.01\pm   0.66$ & $  2.41\pm   0.63$ & $  0.71\pm   0.06$  \\ 
990123 & $143.79\pm  17.78$ & $ 20.28\pm   1.85$ & $  0.88\pm   0.02$  \\ 
990510 & $ 17.63\pm   2.00$ & $ 13.16\pm   1.12$ & $  0.57\pm   0.03$  \\ 
990705 & $ 25.60\pm   2.03$ & $  0.34\pm   0.12$ & $  0.99\pm   0.00$  \\ 
991216 & $ 53.54\pm   5.94$ & $ 36.64\pm   1.79$ & $  0.59\pm   0.03$  \\ 
000210 & $ 16.93\pm   1.41$ & $  0.50\pm   0.12$ & $  0.97\pm   0.01$  \\ 
000926 & $ 27.97\pm   9.90$ & $  9.97\pm   3.75$ & $  0.74\pm   0.10$  \\ 
010222 & $ 85.78\pm   2.17$ & $ 22.79\pm   2.48$ & $  0.79\pm   0.02$  \\ 
011211 & $  6.72\pm   0.86$ & $  1.32\pm   0.22$ & $  0.84\pm   0.03$  \\ 
020405 & $  7.20\pm   0.92$ & $  4.60\pm   1.29$ & $  0.61\pm   0.07$  \\ 
020813 & $ 77.50\pm  31.06$ & $ 22.16\pm   3.15$ & $  0.78\pm   0.07$  \\ 
021004 & $  5.56\pm   0.72$ & $  8.35\pm   1.45$ & $  0.40\pm   0.05$  \\
XRF020903 & $ .0011       $ & $  .01           $ & $  0.10          $  \\ 
\enddata
\tablecomments{The columns are (left to right): (1) GRB name, (2)
radiated energy (3) kinetic energy (4)efficiency.} 
\end{deluxetable}

\begin{figure}
\epsscale{0.5}
\plotone{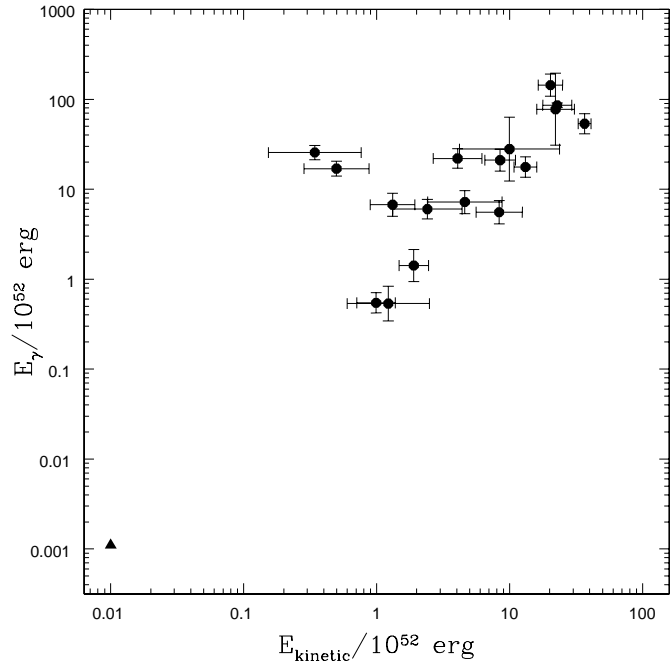}
\caption{Radiated energy vs. kinetic energy for our sample of 17 GRBs.  
The triangular point in
the lower right corner is XRF020903. }
\end{figure}

\begin{figure}
\epsscale{0.5}
\plotone{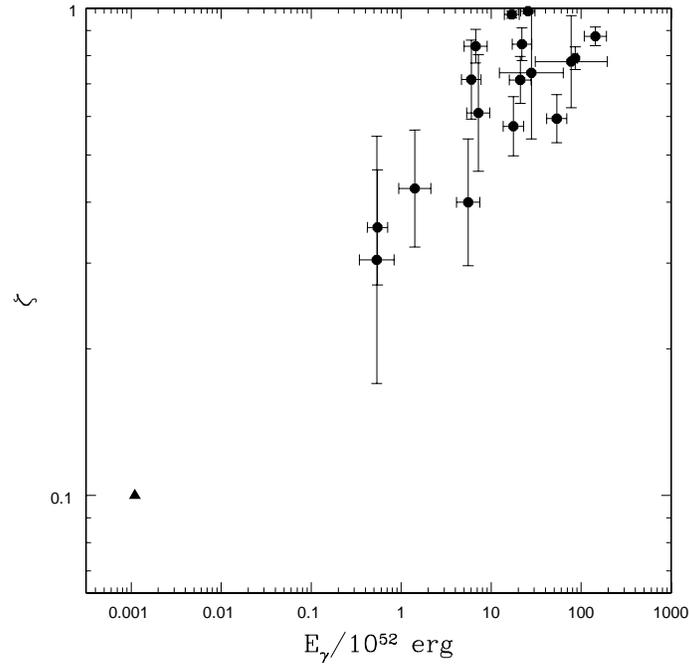}
\caption{Efficiency $\zeta$ vs. radiated energy $E_{\gamma}$. The 
triangular point in the lower right corner is XRF020903.}
\end{figure}

\begin{figure}
\epsscale{0.5}
\plotone{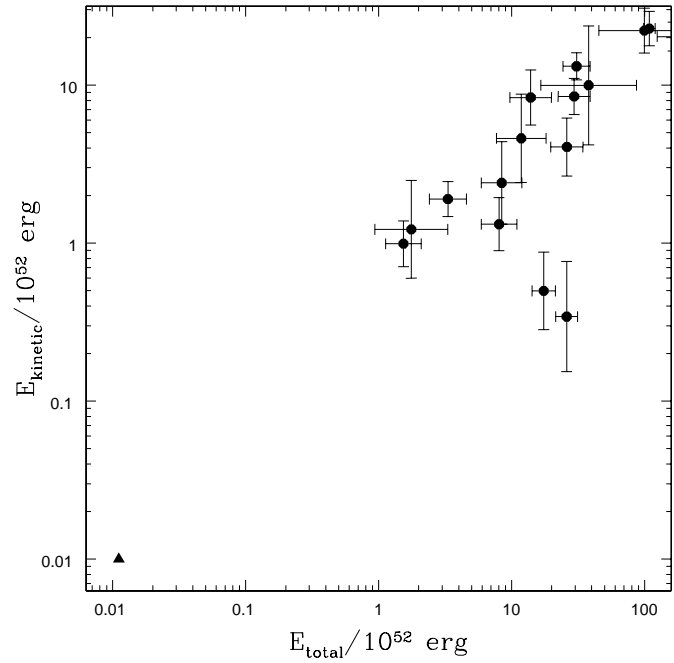}
\caption{Kinetic energy vs total energy.
The triangular point in
the lower right corner is XRF020903. }
\end{figure}

\begin{figure}
\epsscale{0.5}
\plotone{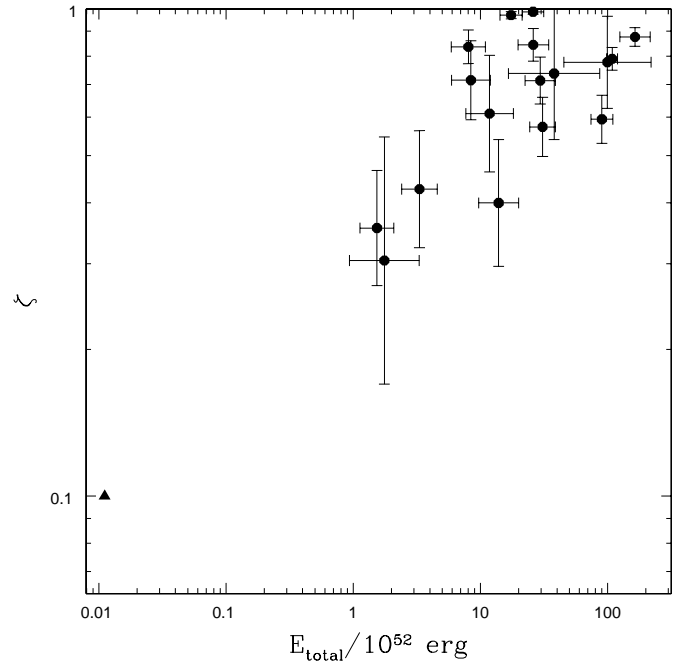}
\caption{Efficiency $\zeta$ vs. total energy $E_{tot}$. The triangular 
point in
the lower right corner is XRF020903.}
\end{figure}

\begin{figure}
\epsscale{0.5}
\plotone{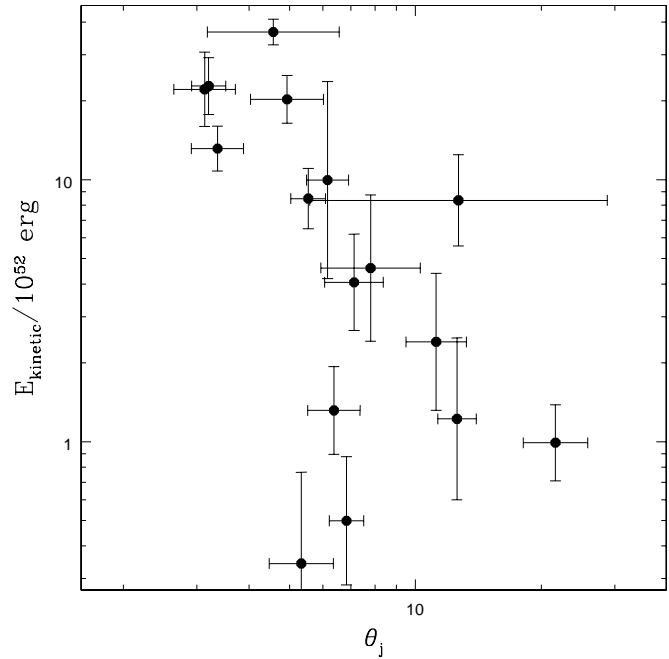}
\caption{Kinetic energy vs jet opening angle. }
\end{figure}

\begin{figure}
\epsscale{0.5}
\plotone{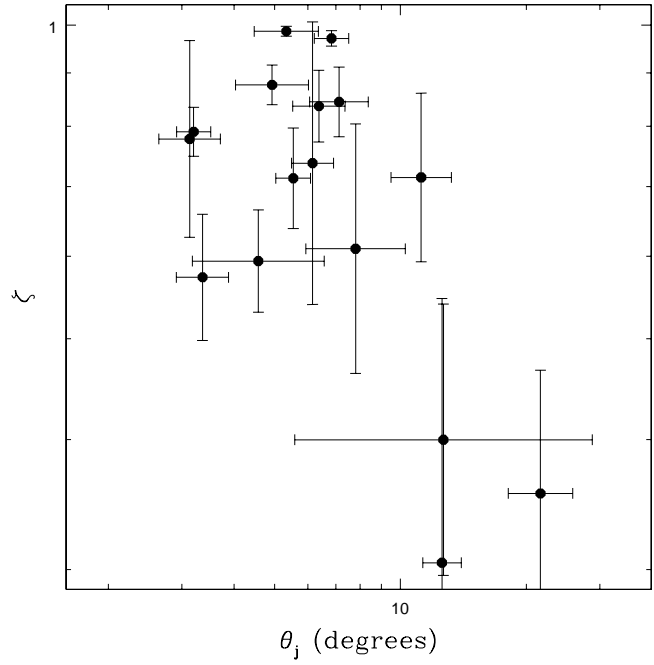}
\caption{Efficiency $\zeta$ vs. jet opening angle $\theta_{j}$.}
\end{figure}

\begin{figure}
\epsscale{0.5}
\plotone{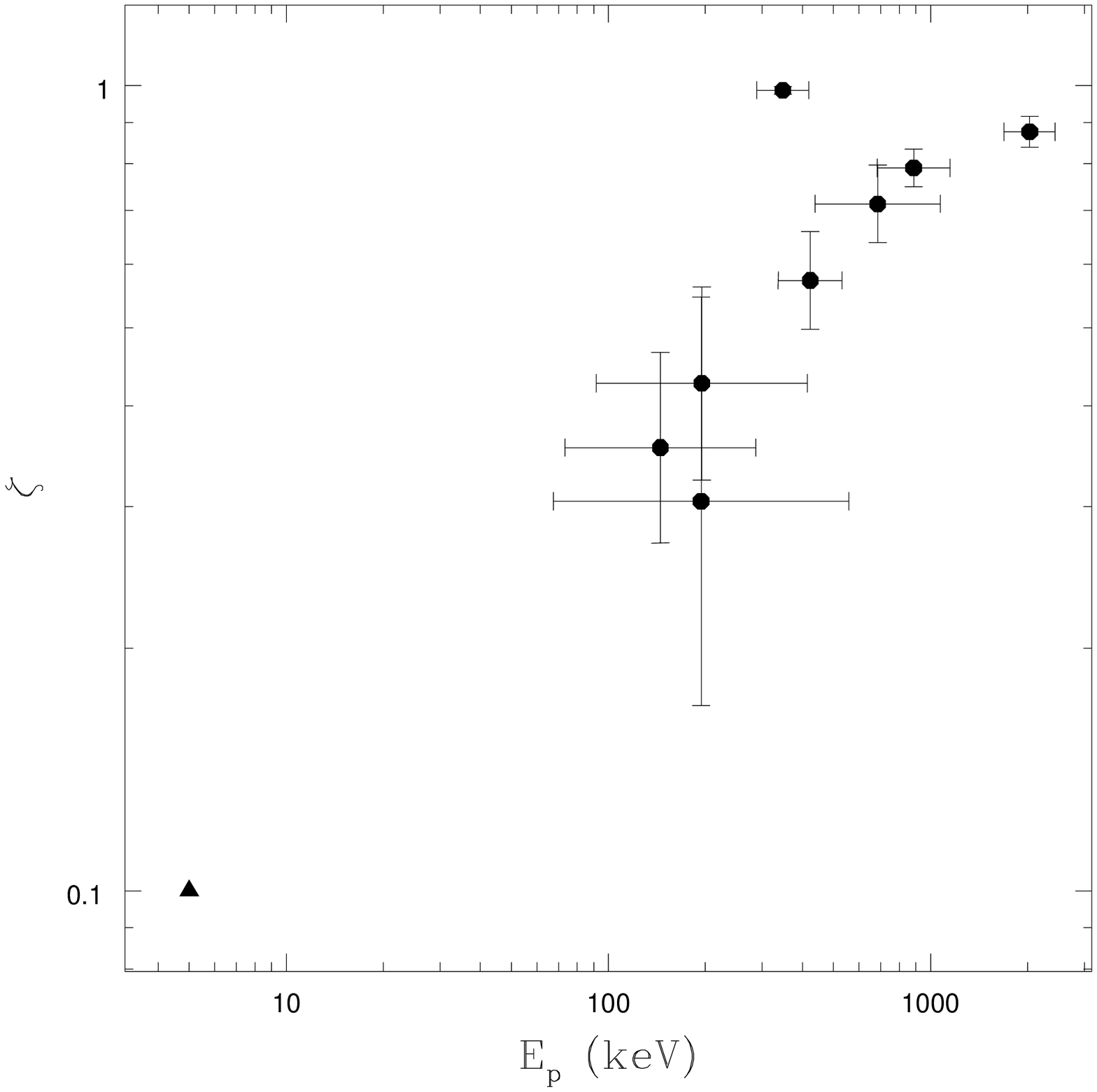}
\caption{Efficiency $\zeta$ vs. spectral peak energy $E_{p}$.
The triangular point in
the lower right corner is XRF020903.}
\end{figure}


\begin{thebibliography}{}
\bibitem[]{}Amati, L., et al. 2002, A\&A, 390, 81 
\bibitem{}Beloborodov, A.M. 2000, ApJ, 539, L25
\bibitem[]{}Berger, E., Kulkarni, S.R,, Frail, D.A. 2003, ApJ, 590, 379  
\bibitem{}Bevington, P.R. \& Robinson, K.D., ``Data Reduction and
Error Analysis for the Physical Sciences'', 2002, McGraw-Hill  
\bibitem{}Bloom, J.S., Frail, D.A., Kulkarni, S.R., 2003, ApJ, 594, 674 
\bibitem{}Fenimore, E. \& Ramirez-Ruiz, E. 2000, astro-ph/0004176
\bibitem{}Frail, D.A., et al 2001, ApJ, 562, 55  
\bibitem{}Freedman,D.L. \& Waxman, E. 2001, ApJ, 547, 922  
\bibitem{}Hurley, K., Sari, R., Djorgovski, S.G. 2002, astro-ph 0211620
\bibitem{} Kobayashi, S. \& Sari, R. 2001, ApJ, 551, 934
\bibitem{} Kumar, P. 1999, ApJ, 532, L113  
\bibitem{}Kumar, P. \& Granot, J. 2003, ApJ, 591, 1075
\bibitem{}Lamb, D.Q., , Donaghy, T. Q., \& Graziani, C. 2004, ApJ,
submitted (astro-ph/0312634)   
\bibitem{}Liang, E.W., Dai, Z.G., Wu, X.F. 2004, ApJL in press, astro-ph 
0403397
\bibitem{}Lloyd, N.M., Petrosian, V.,  Mallozzi, R.S., 2000, ApJ, 534, 
227 
\bibitem{}Lloyd-Ronning, N.M., Dai, X. \& Zhang, B. 2004, ApJ, 601,
371 
\bibitem{}M\'esz\'aros, P. \& Rees, M. J. 1997, 476, 232
\bibitem{}Panaitescu, A., Spada, M. \& Meszaros, P. 1999, ApJ, 522, L105 
\bibitem{}Panaitescu, A. \& Kumar, P. 2001, ApJ, 560, L49  
\bibitem{}Piran, T., Kumar, P., Panaitescu, A. \& Piro, L. 2001, ApJ,
560, L167 
\bibitem{} Press, W.H., et al., ``Numerical Recipes'', 1994, Cambridge 
University Press 
\bibitem{}Reichart, D. et al. 2001, ApJ, 552, 57
\bibitem{} Rossi, E., Lazzati, D., Rees, M.J. 2002, MNRAS, 332, 945
\bibitem{}Sari, R. 1997, ApJ, 489, L37
\bibitem{}Sari, R. , Piran, T. \& Narayan, R. 1998, ApJ, 497, L17 
\bibitem{}Soderberg, A. et al. 2004, ApJ in press, astro-ph 0311050 
\bibitem{}Yost, S., Harrison, F. A., Sari, R. \& Frail, D. A. 2003, ApJ, 
597, 459 
\bibitem{}Zhang, B., Dai, X., Lloyd-Ronning, N. \& M\'esz\'aros,
P. 2004, ApJ, 601, L119 
\bibitem{} Zhang, B. \& M\'esz\'aros, P. 2002, ApJ, 571, 876 
\end{thebibliography}
\end{document}